\documentclass[prb,aps,showpacs,preprint,superscriptaddress]{revtex4}
\usepackage{graphics}
\usepackage{epsfig}
\usepackage{amsmath}
\usepackage{amsfonts}
\usepackage{amssymb}
\usepackage{graphicx}
\usepackage{color}

\begin{document}

\title{Effect of spin orbit coupling and Hubbard $U$ on the electronic structure of IrO$_2$}

\author{S. K. Panda}
\affiliation{Centre for Advanced Materials, Indian Association for the Cultivation of Science, Jadavpur, Kolkata-700032, India}
\author{S. Bhowal}
\affiliation{Department of Solid State Physics, Indian Association for the Cultivation of Science, Jadavpur, Kolkata, 700032, India}
\author{A. Delin}
\affiliation{Department of Materials and Nanophysics, School of Information and Communication Technology,
Electrum 229, Royal Institute of Technology (KTH), SE-16440 Kista, Sweden}
\affiliation{SeRC (Swedish e-Science Research Center), KTH, SE-10044 Stockholm, Sweden}
\affiliation{Department of Physics and Astronomy, Uppsala University, P.O. Box 516, SE-751 20 Uppsala, Sweden}
\author{O. Eriksson}
\affiliation{Department of Physics and Astronomy, Uppsala University, P.O. Box 516, SE-751 20 Uppsala, Sweden}
\author{I. Dasgupta}
\email{sspid@iacs.res.in}
\affiliation{Centre for Advanced Materials, Indian Association for the Cultivation of Science, Jadavpur, Kolkata-700032, India}
\affiliation{Department of Solid State Physics, Indian Association for the Cultivation of Science, Jadavpur, Kolkata, 700032, India}
\begin{abstract}
We have studied in detail the electronic structure of IrO$_2$ including spin-orbit coupling (SOC) and electron-electron interaction, both within the GGA+U and GGA+DMFT approximations. Our calculations reveal that the Ir t$_{2g}$ states at the Fermi level largely retain the J$_{\rm eff}$ = $\frac{1}{2}$ character, suggesting that this complex spin-orbit entangled state may be robust even in metallic IrO$_2$. We have calculated the phase diagram for the ground state of IrO$_2$ as a function of $U$ and find a metal insulator transition
that coincides with a magnetic phase change, where the effect of SOC is only to reduce the critical values of $U$
necessary for the transition.  We also find that dynamic correlations, as given by the GGA+DMFT calculations, tend to suppress the spin-splitting, yielding a Pauli paramagnetic metal for moderate values of the Hubbard $U$.  Our calculated optical spectra and photoemission spectra including SOC are in good agreement with experiment demonstrating the importance of SOC in IrO$_2$. 
\end{abstract}

\pacs{71.20.-b,71.30.+h}

\maketitle
\section{Introduction}
In recent years, $5d$ based oxides have attracted considerable attention where a combined influence of band-structure, electron
correlation, and spin orbit coupling lead to emergent quantum
phenomena.~\cite{JHalf_Kim,Ba2IrO4_Insulator,Na2IrO3,MIT_Dimension,QSL_Na4Ir3O8,QSL_Ba3IrTi2O9,TI_Na2IrO3_1,TI_Na2IrO3_2} 
Until a few years ago, the common belief has been that due to the extended nature of the $5d$ orbitals, the ratio between effective electron-electron
interaction and bandwidth, $U$/$W$ (where $U$ is the Coulomb interaction and $W$ is the bandwidth) is quite
small in $5d$ transition metal oxides (TMO) and density functional theory (DFT) within local density approximation (LDA) or
generalized gradient approximation (GGA) can explain the metallic ground state of these systems. 
Contrary to this expectation there are recent reports of an insulating 
antiferromagnetic ground state in $5d$ TMO, e.g. Sr$_2$IrO$_4$, Ba$_2$IrO$_4$, and Na$_2$IrO$_3$,~\cite{Sr2IrO4_Insulator,Ba2IrO4_Insulator,Na2IrO3}
where in addition
to the crystal field and Coulomb repulsion strong spin orbit coupling plays a key role. 
In d$^5$ Ir oxides, due to large crystal field splitting 
and strong SOC the t$_{2g}$ orbitals are renormalized into doubly degenerate J$_{\rm eff}$ = 1/2 and quadruply
degenerate J$_{\rm eff}$ = 3/2 states, leading to a narrow band of half filled J$_{\rm eff}$ = 1/2 states.~\cite{JHalf_Kim} 
Inclusion of moderate Coulomb interaction in the spin orbit entangled J$_{\rm eff}$ = 1/2 manifold opens up a gap explaining the insulating
property of some of the iridates.~\cite{JHalf_Kim}
Very recently it has been speculated that in contrast to the orbitally ordered states
in $3d$ insulating oxides, the spin-orbital entangled J$_{\rm eff}$ = 1/2 state is robust and does not melt 
away even in itinerant metallic systems.~\cite{JeffHalf_IrO2} This possibility was recently suggested for metallic
IrO$_2$ using resonant x-ray diffraction experiments~\cite{JeffHalf_IrO2} where it was argued that Ir $5d$ t$_{2g}$ 
orbitals at the Fermi level are fairly close to the J$_{\rm eff}$ = 1/2 state due to strong spin orbit coupling. 
The importance of spin orbit coupling in IrO$_2$ is also manifested by the recent observation of large
spin Hall effect~\cite{SpinHallEffect_IrO2} which is a novel topological transport phenomena caused by spin-orbit interaction. 
\begin{figure}
\centering
\includegraphics[width=\columnwidth]{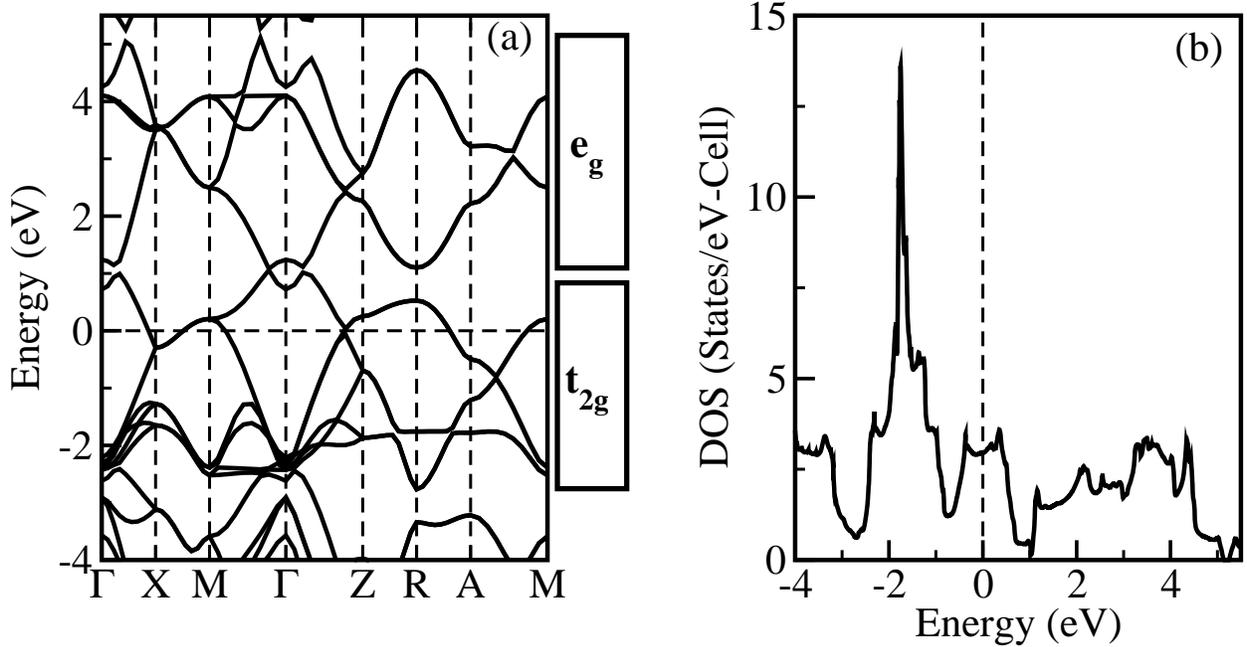}
\caption{Non spin polarized (a) band dispersion along various high symmetry directions and (b) density of states, computed within the GGA approximation.}
\label{dos_band_nm_gga}
\end{figure}   
\par
IrO$_2$ crystallizes in the rutile type structure with two formula units per unit cell. The electronic structure
as well as optical properties of IrO$_2$ have been investigated by several
groups~\cite{IrO2_th_Mattheiss,IrO2_th_Xu,IrO2_th_Ahuja} in the past but none
of these calculations analyzed the possibility of the J$_{\rm eff}$ = 1/2 state in this system. 
Further, there are no studies where the combined role of spin orbit interaction and Coulomb
correlation is analyzed in detail. In the present paper, we have investigated
the electronic structure of IrO$_2$ using density functional theory (DFT) in the framework
of GGA+SOC+Hubbard $U$ (GGA+SOC+U) as well as GGA+SOC+dynamical mean field theory (GGA+SOC+DMFT) calculations. 
Our GGA+SOC+U calculations as a function of $U$ reveal that nonmagnetic metallic IrO$_2$ transforms
to an antiferromagnetic metal and eventually into an antiferromagnetic Slater insulator. 
The GGA+SOC+DMFT calculations result in a suppressed exchange splitting, for moderate values of the Hubbard $U$.
We have analyzed the nonmagnetic metallic state in detail and examined
the suggestions for the J$_{\rm eff}$ = 1/2 state. In addition we have also calculated the
optical conductivity and the photoemission spectra including SOC and Coulomb correlation
and compared with available
experiments. 
The remainder of the paper is organized as follows. In Sec. II, we discuss
the crystal structure and the computational details. Section III is devoted to
results and discussions followed by conclusions in Sec. IV. 
\begin{figure}
\includegraphics[width=\columnwidth]{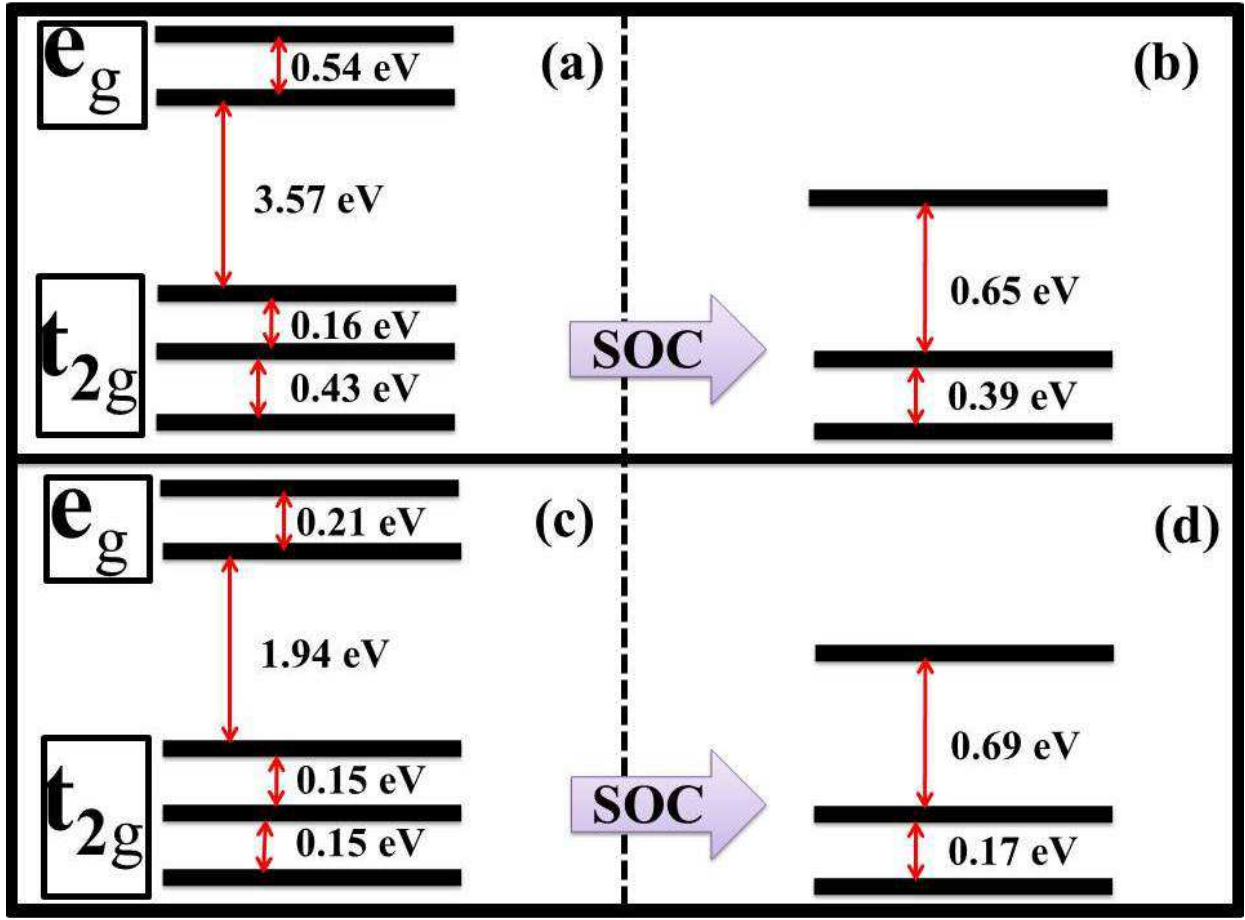}
\caption{(Color online) Crystal field splitting, obtained from NMTO calculation for (a) experimental and (c) ideal structure. Effect of SOC
on the t$_{2g}$ states of (b) experimental structure and (d) ideal structure.}
\label{crystalfield}
\end{figure} 
\section{Computational detail and Crystal Structure}
The density functional theory calculations have been performed using three
different methods, namely (a) the plane wave based method as implemented
in the Vienna \textit{ab initio} simulation package (VASP),~\cite{vasp1,vasp2}
(b) the full potential linearized augmented plane wave (FP-LAPW) method~\cite{wien2k} and (c) the full potential linearized muffin-tin orbital (FP-LMTO) method.~\cite{FPLMTOCode} We have checked that all the three methods yield essentially identical band structures for IrO$_2$.
In order to find out the ground state,
the plane wave calculations were performed within the local (spin) density approximation (LSDA), with generalized
gradient correction (GGA) of Perdew-Burke-Ernzerhof, with and without including
Hubbard $U$~\cite{LDAU_Liech} and SOC.~\cite{VASP_SOC} The kinetic energy cutoff
of the plane wave basis was chosen to be $600$~eV. 
Brillouin-zone integration have been performed using a $14\times 14\times 20$ k-mesh. For the calculation of the optical spectra
corresponding to the nonmagnetic metallic state, we have employed the all-electron
FP-LAPW method. The muffin-tin radii ($R_{MT}$) of Ir and O are chosen to be
1.09~\r{A}, and 0.87~\r{A}, respectively. 
To achieve energy convergence of the eigenvalues,
the wave functions in the interstitial region were expanded in plane waves with a cutoff $R_{MT}k_{max}$=7, where $R_{MT}$ 
denotes the smallest atomic sphere radius and $k_{max}$ represents the magnitude
of the largest k vector in the plane wave expansion. The valence wave functions inside the spheres are expanded up
to $l_{max}$=10, while the charge density is Fourier
expanded up to $G_{max}$=12. 
\par
All the GGA+DMFT calculations have been carried out using a full potential linear muffin-tin orbital (FP-LMTO) method~\cite{FPLMTOCode}
as implemented in the RSPT code. In this implementation of GGA + DMFT the many-body corrections appear in a form which depends on a self-consistently calculated density matrix and on the correlated orbitals.~\cite{FPLMTOCode,DMFT_transitionMetal1FLEX} In the present case  the  correlated orbitals are 5d states on the Ir atoms. Hence, the calculations treat in equal footing spin-orbit effects, crystal field splittings, band formation as well as electron-correlations. The effective impurity problem in the GGA+DMFT calculations has been solved through the spin polarized T-matrix fluctuation-exchange (SPTF) solver.~\cite{FLEX2} The SPTF solver has been chosen as it is known to be very efficient for moderately correlated systems ( U $\leq$ W) and has been successfully applied to various materials.~\cite{DMFT_transitionMetal1FLEX,FLEX1,FLEX2,DMFT_transitionMetal3} The SPTF solver is based on a perturbation expansion in the Coulomb interaction, where the Hubbard $U$ is considered to be smaller than the band-width. In the past it has been used with success for heavy elements~\cite{FLEX2} where, as for IrO$_2$, the spin-orbit effects are important. 
\begin{figure}
\centering
\includegraphics[width=\columnwidth]{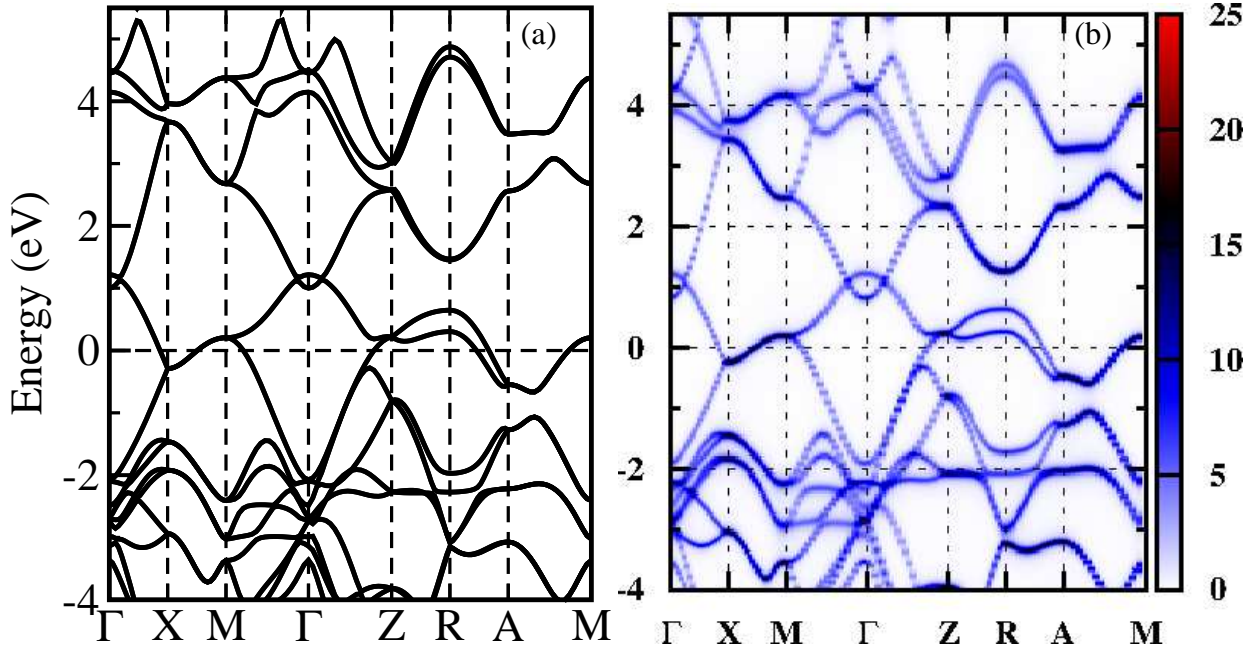}
\caption{(Color online) (a) Band dispersion computed within GGA+U+SOC approximation and (b) the k-resolved total spectral function A(k,$\omega$) along the high-symmetry directions of the Brillouin zone, computed within GGA+DMFT+SOC approximation. The vertical color scale shows the intensity of A(k,$\omega$).}
\label{dos_band_nm_ggausoc}
\end{figure}
\par
IrO$_2$ crystallizes within a tetragonal rutile structure, having space group P4$_2$/mnm.
Each unit cell contains two Ir ions, at (0, 0, 0) and ($\frac{1}{2}$, $\frac{1}{2}$, $\frac{1}{2}$) and four O ions
with coordinates (u, -u, 0), (-u, u, 0), ($\frac{1}{2}$-u, $\frac{1}{2}$-u, $\frac{1}{2}$), and (-$\frac{1}{2}$+u, -$\frac{1}{2}$+u, -$\frac{1}{2}$),
where u = 0.3077.~\cite{cryststruct_IrO2} 
Each Ir ion is surrounded by six O ions in a distorted octahedron environment. 
Neighboring IrO$_6$ octahedral units share edges along the c axis and vertices
in the basal plane. 
Each Ir atom has two O atoms as neighbors at a distance d$_1$ = $\sqrt(2)$ua and four at d$_2$ = [2($\frac{1}{2}$-u)$^2$ + ($\frac{c}{2a}$)$^2$]$^{1/2}$a. All the 
Ir-O bond lengths are equal if the parameter u has the value u$^*$ = $\frac{1}{4}$[1+$\frac{1}{2}$($\frac{c}{a}$)$^2$]. The octahedral coordination
of each Ir atom is ideal if ($\frac{c}{a}$)$_{ideal}$ = 2-$\sqrt{2}$ and u$_{ideal}$ = $\frac{1}{2}$(2-$\sqrt{2}$). For IrO$_2$, u$^{*}$ = 0.312 and therefore
u $<$ u$^{*}$.  
All the calculations have been carried out with the experimental structure and the
antiferromagnetic ordering has been simulated by considering an anti parallel alignment of the spin of two Ir ions in the unit cell. 
In addition, we have also carried out electronic structure calculations for the ideal structure in order to assess the impact of
distortion on the electronic structure.
\section{Results and Discussion}
To begin with we have analyzed the non-spin polarized band structure and DOS of rutile IrO$_2$ obtained using the GGA method.
The results of our calculations are presented in Fig.~\ref{dos_band_nm_gga}. Both the DOS and the band structure are in good agreement with the
earlier calculation~\cite{IrO2_th_Ahuja}
on the same system. As discussed earlier in a rutile structure each Ir atom is surrounded by a nearly octahedral array of six O atoms
and the site symmetry of the Ir atom may be considered as a sum of a large octahedral term plus a small orthorhombic distortion.~\cite{IrO2_th_Mattheiss}
In such a crystal field the Ir $d$ orbitals split into three fold degenerate t$_{2g}$ states and two fold degenerate e$_g$ states. These orbitals for the rutile structure are  a linear combination of the $d$-orbitals expressed along the crystallographic a, b, and c axes,~\cite{IrO2_th_Mattheiss} as the O octahedron around each Ir is not aligned along the crystallographic a, b, and c axes. 
The degeneracy of the t$_{2g}$ and e$_g$ orbitals is, however, lifted by the orthorhombic term. As a consequence, the Fermi level
is dominated by six Ir t$_{2g}$ states arising from the two Ir atoms in the unit cell. The e$_g$ states are completely empty and lie
above the Fermi level. The twelve
O $p$ states are below the Ir t$_{2g}$ manifold where again the degeneracy of the O-$p$ states is lifted by the orthorhombic term.
\par
Next we have calculated
the crystal field splitting at the Ir $d$ site. For this purpose, the N-th order muffin tin orbital (NMTO) downfolding calculations~\cite{nmto} 
were carried out keeping only the Ir $d$ states
in the basis and downfolding the O-$p$ states. The onsite block of the real space Hamiltonian provide the crystal field splitting 
at the Ir site where the O covalency effect is also taken into account. The crystal field splitting of the Ir-$d$ state for the experimental structure is shown
in Fig.~\ref{crystalfield} (a) and is consistent with the D$_{2h}$ symmetry of the Ir site where the degeneracy of all the $d$ orbitals is completely removed. 
Our calculations reveal that the e$_g$ block is separated from the t$_{2g}$ complex by 3.6 eV. Since the e$_g$ block is completely
empty, we shall concentrate on the t$_{2g}$ block and the crystal field term for the t$_{2g}$ block may be written as 
$$
\quad
\begin{pmatrix}
 -2\epsilon & 0 & 0 \\
 0 & \epsilon & t \\
0 & t & \epsilon 
\end{pmatrix}
\quad
$$
where we obtain $\epsilon$ = 0.17 eV and t=0.08 eV from our NMTO calculation.
In order to assess the role of distortion on the crystal field splitting, we have also carried out a GGA calculation for the ideal structure of
IrO$_2$ with a = 5.3919 \AA{}, c = 3.1586 \AA{} and u = 0.2929. While we find substantial splitting between the e$_g$
and the t$_{2g}$ states, the intra-t$_{2g}$ splitting is now appreciably reduced and the crystal field term for the t$_{2g}$ block
is calculated to be $\epsilon$ = $t$ = 0.075 eV.
The details of the crystal field splitting are shown in Fig.~\ref{crystalfield} (c). 
\begin{figure}
\centering
\includegraphics[width=\columnwidth]{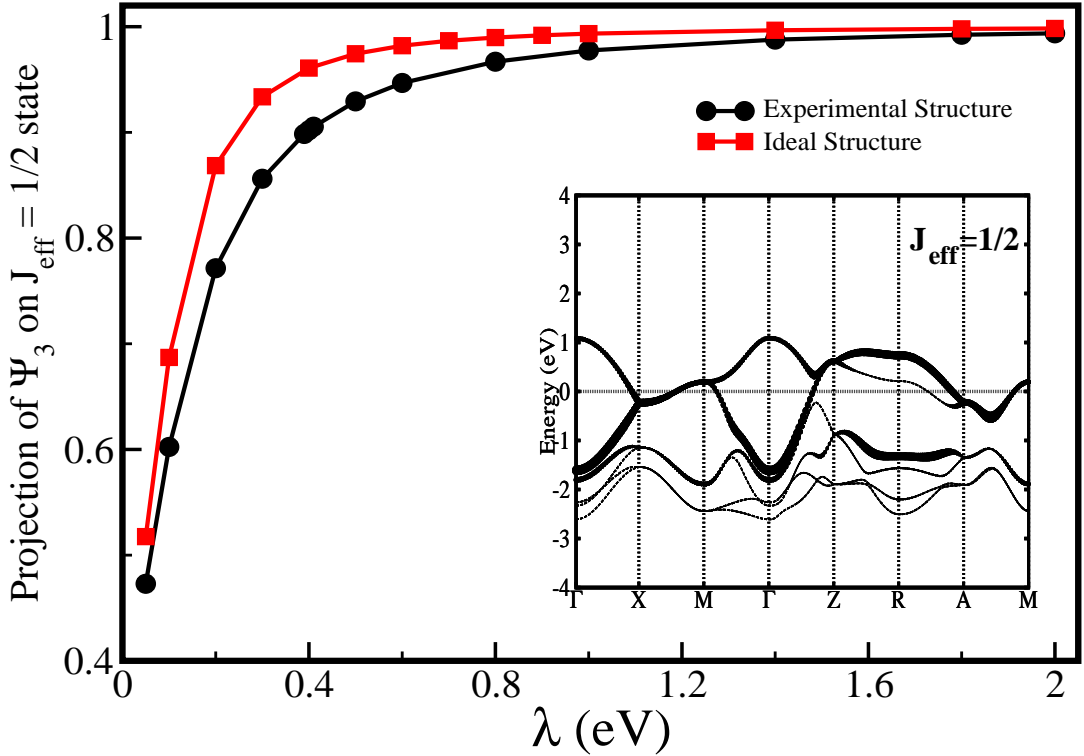}
\caption{(Color online) Projection of $|\psi^{\pm}_3\rangle$ on the J$_{\rm eff}$ = 1/2 state. The inset shows the low energy bands for the t$_{2g}$ states decorated with J$_{\rm eff}$ = 1/2 character.}
\label{projection_jhalf}
\end{figure}

\begin{figure}
\centering
\includegraphics[width=\columnwidth]{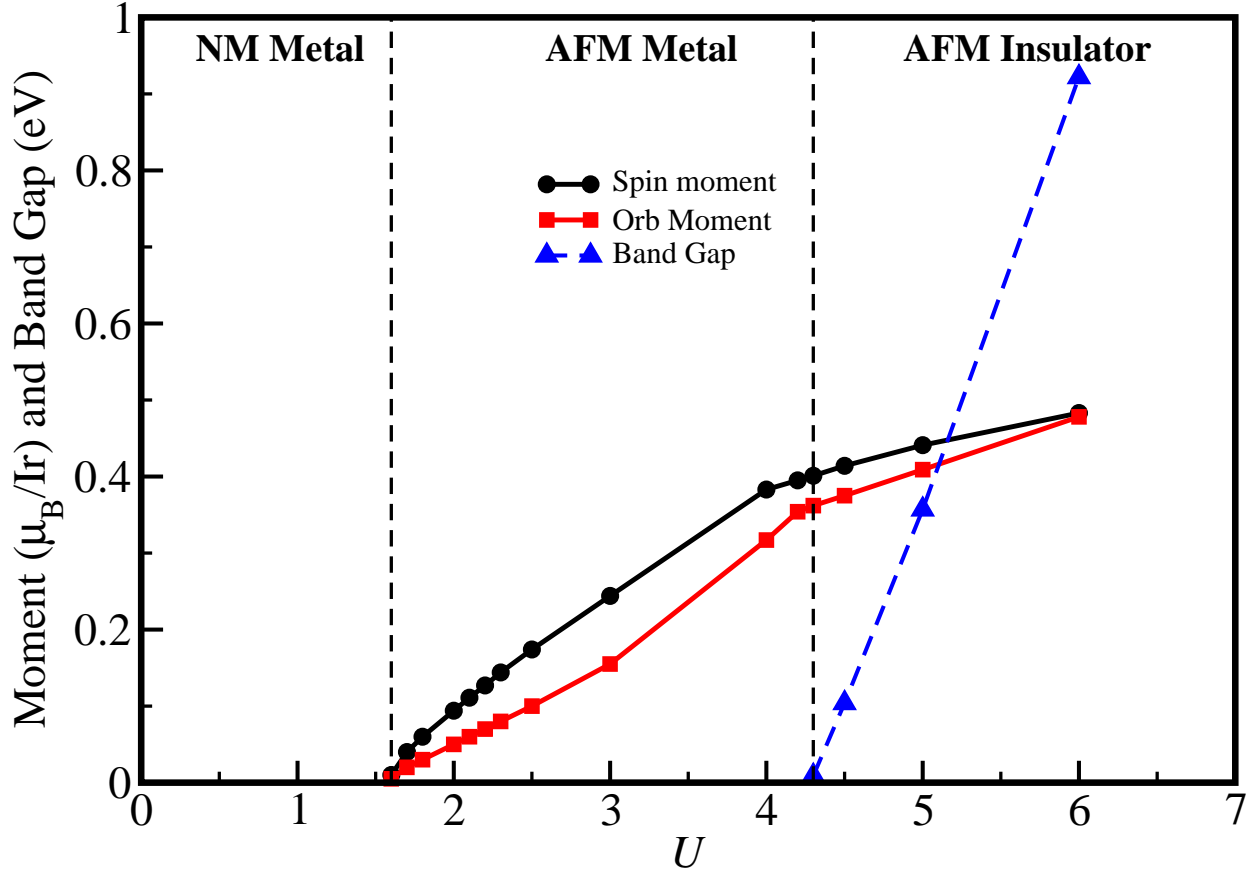}
\caption{(Color online) The phase diagram for the ground state of IrO$_2$ as a function of $U$ within GGA+U+SOC approach.
Change of the spin and orbital moment per Ir site and the band gap in the antiferromagnetic phase as a function of $U$ are also shown.}
\label{bandgap}
\end{figure}  
\par
We have next considered the spin orbit interaction in our calculations, where the magnetization direction was chosen along (001). In addition to spin-orbit coupling, we have also included a Hubbard $U$ = 2 eV
in order to consider the effect of electron correlations in IrO$_2$. 
These calculations were done on the GGA+U level, as well as employing dynamical mean field theory. Recently the electronic structures of several iridium based oxides have been investigated both in the framework of GGA+U+SOC~\cite{JHalf_Kim,Mazin,Subedi,Cao} and GGA+DMFT+SOC.~\cite{Martins,Zhang}
The result of our calculation for IrO$_2$ is displayed in Fig.~\ref{dos_band_nm_ggausoc}.
The spin orbit coupling (SOC) leads to important changes in the band structure in the t$_{2g}$ manifold near the Fermi level. As a result of SOC the t$_{2g}$ states are further split in such a way that the Fermi level now lies on a pair of bands separated from the rest. Spin-orbit effects are in general dependent on the degree of hybridization as well as on the symmetry of the eigenstates, which varies across the Brillouin zone.~\cite{IrO2_th_Mattheiss} In particular, along the direction ZRA the degeneracy of the bands is removed in the relativistic limit where  spin-orbit coupling induces a splitting of approximately 0.5 eV. As a consequence of SOC, the t$_{2g}$ states are grouped in such a way that there is a pair of bands forming a quartet. These bands are fully occupied and are close to each other while the Fermi level is on a band forming a doublet. The former is reminiscent of J$_{\rm eff}$ = 3/2 states and the latter of J$_{\rm eff}$ = 1/2 states that have been discussed in the literature to understand the physics of iridates.~\cite{JHalf_Kim} Recently x-ray absorption spectroscopy emphasized the importance of $j$ quantum states due to strong SOC in IrO$_2$.~\cite{5dSOC_JPhys}
The division of the t$_{2g}$ orbitals with an effective quantum state L$_{\rm eff}$ = 1 forming a J$_{\rm eff}$ = 3/2 quartet and a J$_{\rm eff}$ = 1/2 doublet
not only requires large SOC but also completely degenerate t$_{2g}$ states that are well separated from the e$_g$ states. 
Any kind of mixing between the t$_{2g}$ and e$_g$ states or the breaking of the three-fold degeneracy of the t$_{2g}$ states will lead to a deviation
from the J$_{\rm eff}$ = 1/2 state. As the orthorhombic distortion lifts the degeneracy of the t$_{2g}$ states, the existence of the J$_{\rm eff}$ = 1/2
state even in metallic IrO$_2$ as suggested by the resonant x-ray diffraction experiment~\cite{JeffHalf_IrO2} therefore requires further scrutiny.
\par
In view of the above, we have examined the validity of the 
J$_{\rm eff}$ = 1/2 state in IrO$_2$ based on a model Hamiltonian with realistic crystal field splitting and SOC. We have first considered the onsite term of the down-folded Hamiltonian for the t$_{2g}$ block in the presence of spin orbit coupling which may be written as
\begin{align}
H &= \begin{pmatrix}
    H_+ & 0 \\
    0 & H_-
\end{pmatrix}
\end{align}
where
\begin{align}
H_{\pm} &= \begin{pmatrix}
    -2\epsilon & \pm\frac{\lambda}{2} & -i\frac{\lambda}{2}  \\
    \pm\frac{\lambda}{2} & \epsilon & t\mp i\frac{\lambda}{2} \\
     i\frac{\lambda}{2} & t\pm i\frac{\lambda}{2} & \epsilon
\end{pmatrix}
\end{align}
here for the representation of $H_{+}$ and $H_{-}$ we have employed the basis functions ( $|xy +\rangle$, $|yz -\rangle$, $|zx -\rangle$) and ($|xy -\rangle$, $|yz +\rangle$, $|zx +\rangle$), respectively. The  
spin orbit interaction is represented by the parameter $\lambda$. Each eigenstate of H$_{+}$ has its counterpart in an eigenstate of H$_{-}$ for the Kramer's doublet. 
The highest state becomes the J$_{\rm eff}$ = 1/2 state given by
\begin{equation}
\psi^{\pm}_{J_{\rm eff}=1/2} = \frac{1}{\sqrt{3}}[|xy \pm\rangle + |yz \mp\rangle + i|zx \mp\rangle]. 
\end{equation}
\par
In Fig.~\ref{crystalfield} (b) and (d), we show the Ir t$_{2g}$ levels in the presence of spin-orbit coupling both for the experimental structure
and the idealized structure for $\lambda$ = 0.5 eV, a value typical for the iridates.~\cite{ClancyXAS}
Fig.~\ref{crystalfield} (b) and (d) clearly reveal that the t$_{2g}$ levels are renormalized upon spin orbit coupling. 
Next we have calculated the projection $\mid\langle\psi^{\pm}_{J_{\rm eff}=1/2}|\psi^{\pm}_3\rangle\mid^2$
where $|\psi^{\pm}_3\rangle$ is the eigenstate corresponding to the highest eigenvalue of the Hamiltonian H$_{\pm}$ (see equation (2)).
The results of our calculation for the experimental as well as for the idealized structure are shown
as a function of $\lambda$, in Fig.~\ref{projection_jhalf}.
We gather from Fig.~\ref{projection_jhalf} that the J$_{\rm eff}$ = 1/2 character depend crucially on the strength of spin orbit coupling $\lambda$ and $|\psi^{\pm}_3\rangle$ has about 90$\%$ J$_{\rm eff}$ = 1/2 character for the experimental structure and about 96$\%$ for the idealized structure for $\lambda$ = 0.5 eV. It is interesting to note that the J$_{\rm eff}$ = 1/2 character largely survives even for non-degenerate t$_{2g}$ states in the presence of strong spin-orbit coupling.  As the bare width of the t$_{2g}$ state in IrO$_{2}$ is larger than the SOC ($\lambda \sim$~0.5~eV), the J$_{\rm eff}$ = 3/2 levels are mixed with the J$_{\rm eff}$ = 1/2 ones and the highest Kramer's doublet, may deviate from the pure J$_{\rm eff}$ = 1/2 character. In order to clarify that, in addition to the analysis based on the on-site term as described above, we have derived a low energy tight-binding model including spin orbit coupling for the t$_{2g}$ states of IrO$_{2}$ where the  hopping parameters are obtained from our NMTO downfolding calculations. The projection of the J$_{\rm eff}$ = 1/2 character on the tight binding band structure (See Fig.~\ref{projection_jhalf}(inset)) clearly reveal that there is small  hybridization between J$_{\rm eff}$ = 1/2 and J$_{\rm eff}$ = 3/2 bands in some regions of the Brillouin zone but importantly the pair of bands at the Fermi level largely retain the J$_{\rm eff}$ = 1/2 character and it does not melt away in the metallic state as suggested in Ref.~\onlinecite{JeffHalf_IrO2}, indicating the robustness of the J$_{\rm eff}$ = 1/2 state for large enough values of SOC.
\par
In order to check the reliability of the GGA+U+SOC calculations for metallic IrO$_2$, we have also carried
out GGA+DMFT+SOC calculations and the resulting k-resolved spectral density is shown in Fig.~\ref{dos_band_nm_ggausoc} (b). We find 
that the quasiparticle features are protected and the basic structure of the k-resolved spectral density is very similar to the GGA+U+SOC
 calculations (Fig.~\ref{dos_band_nm_ggausoc} (a)).
\begin{figure}
\centering
\includegraphics[width=\columnwidth]{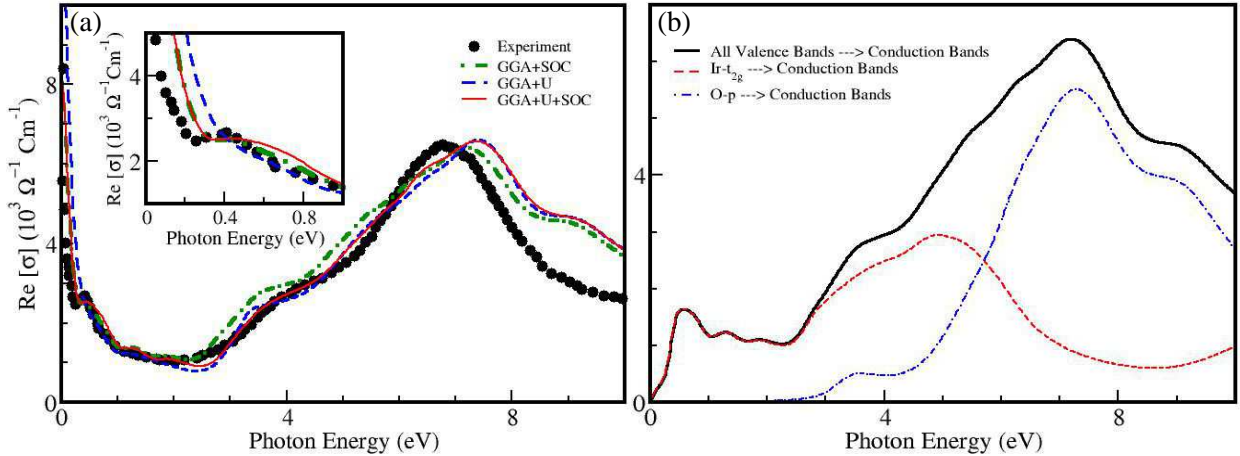}
\caption{(Color online) (a) A comparison of the experimental and theoretical optical conductivity as discussed in the text. Experimental data are taken from Ref.~\onlinecite{Optics_Exp}. The inset shows that the small peak around 0.4~eV is only captured when SOC is included. (b) Interband part of the optical conductivity.}
\label{optics}
\end{figure} 
\par
Next we have investigated whether IrO$_2$ can be made magnetic as well as  insulating upon increasing the value of $U$.
In practical experimental situations, this can be done by tuning the ratio between $U$ and the bandwidth, $W$. This can be done experimentally in several ways, e.g. by means of alloying that narrows the bandwidth either by reducing the direct overlap between the Ir $5d$ orbitals, or by means of negative chemical pressure. Furthermore, the relative importance of $U$ can be enhanced by reducing the t$_{2g}$ bandwidth by 
intercalation.
The results of our GGA+U+SOC calculations are shown in Fig.~\ref{bandgap}, where we plot
the magnetic moment and the band gap. Fig.~\ref{bandgap} clearly establishes a phase diagram for the ground state of IrO$_2$,
showing that it is essentially a non-magnetic
metal for $U<1.6$~eV, an antiferromagnetic metal for 1.6~eV~$\leq$~$U$~$<$~4.3~eV and an antiferromagnetic insulator for $U$ $\geq$~4.3~eV.
As expected, the magnetic moment increases monotonously and the band gap increases nearly linearly
with $U$, beyond two critical values of $U$, $U_{c1} \simeq$ 1.6 eV and $U_{c2} \simeq$ 4.3 eV for the magnetic transition and
the metal-insulator transition, respectively.
It is interesting to note that SOC is neither necessary for stabilizing magnetism nor for the existence of metal insulator transition. The role of SOC is merely
to reduce the critical value of $U_{c1}$ and $U_{c2}$. The ratio $\left <l_z \right >/\left <m_z\right >$ (where $\left <m_z\right > = 2\left <s_z\right >$) deviates from the ideal value of 2 due to the deviation
from the J$_{\rm eff}$ = 1/2 character. However for the ideal structure the J$_{\rm eff}$ = 1/2 behavior is restored.
\par
The data in Fig.~\ref{bandgap} show that already for a rather modest value of $U$, one finds a non-magnetic state. For instance, in the GGA+U calculation, a value of $U$ = 2 eV results in a spin-polarized ground state, and one may ask how accurate GGA+U calculation is in reproducing the ground state properties of this material, especially since a value of $U$=1.5-2.0 eV is rather realistic for the $5d$ orbitals in IrO$_2$. Experimentally it is known that IrO$_2$ is a Pauli paramagnet, with a spin-degenerate ground state configuration, and the data in Fig.~\ref{bandgap} suggest that only $U$ values less than 1.6 eV are consistent with this experimental fact. To further analyze this situation we carried out spin-polarized DMFT calculations for the case when $U$=2 eV. In this case we found that a spin-degenerate solution is stable, and hence that the experimental ground state is reproduced for realistic values of $U$. The dynamic correlations embodied within GGA+DMFT calculation tend to suppress the exchange splitting for moderate values of the Hubbard $U$. It is expected that the DMFT results would give qualitatively the same results as those shown in Fig.~\ref{bandgap}, albeit with slightly large values of the critical $U$-values for the transition to the AFM metal and AFM insulating phase. This was not pursued further here. 
\begin{figure}
\centering
\includegraphics[width=\columnwidth]{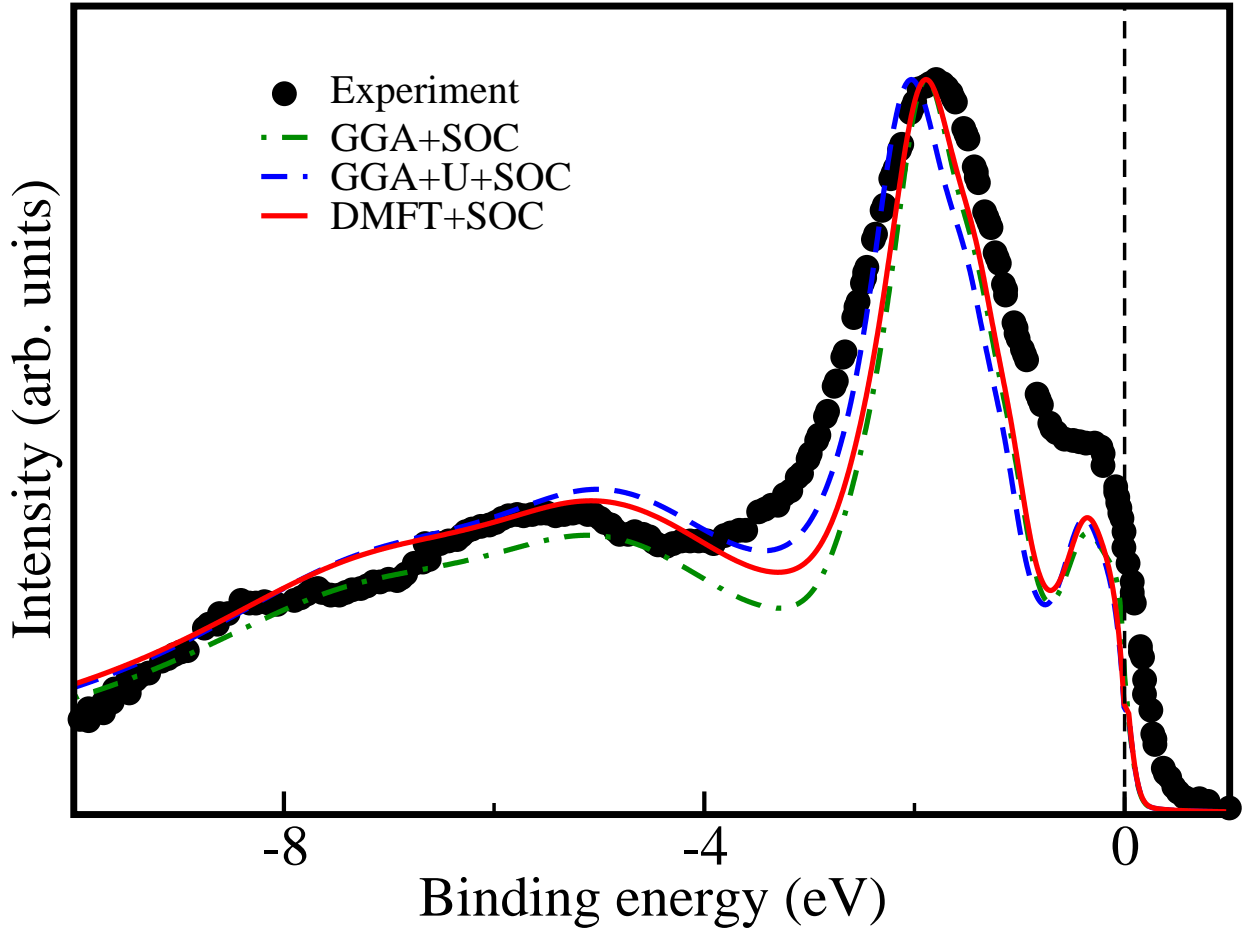}
\caption{(Color online) A comparison of the PES. The experimental data has been taken from Ref.~\onlinecite{IrO2_PES}.}
\label{dmft_spectra}
\end{figure} 
\par
Finally we have calculated the optical properties and the photoemission spectra of IrO$_2$ including the effect of SOC. 
The optical conductivity
has two components because of the tetragonal symmetry of the bravais lattice, one parallel and the other perpendicular to the c-axis. 
In order to compare with experimental results on polycrystalline samples,~\cite{Optics_Exp} we calculate the  average of the two theoretically computed
interband components of the optical conductivity as
$[2\times \sigma_{xx}(\omega) + \sigma_{zz}(\omega)]/3$. 
The interband part of the optical conductivity is broadened by a Lorentzian function.
The intraband Drude component of the optical conductivity
is calculated from the bare plasma frequencies and has been added to each interband part separately to obtain the full optical conductivity.
Fig.~\ref{optics} (a) displays the experimental optical conductivity along with the computed optical conductivity with and without SOC. 
The small peak seen around 0.4~eV is only captured when SOC is included (see inset of Fig.~\ref{optics} (a)). The amplitude of the real part of the optical conductivity in the entire energy range is sightly different when SOC is included. The calculated spectra within GGA+U+SOC approach shows a feature-for-feature similarity when compared with the
experimental results except that the main experimental peak, located around 6.5 eV
appear at higher photon energy in our calculation. 
Similar disagreement has also been reported for fcc Ni~\cite{Optics_high} and for some of the spectral properties of CeN.~\cite{Optics_high1} The fact that the calculated peak appears slightly higher than experiment may indicate that a non-Hermitian energy dependent self-energy is important for the correct description of the optical properties. This has been shown to be the case for other moderately correlated systems.~\cite{Optics_Perlov} To understand the origin of this high energy peak, we have computed the optical conductivity corresponding to interband transitions from the Ir $t_{2g}$ valence states to all the conduction states and also O $p$ valence states to the conduction states. Our results, as presented in Fig.~\ref{optics} (b) clearly shows that the main peak arise due to interband transition from O $p$ states to the conduction states. 
\par
The calculated DOS is usually compared with the photoemission spectra (PES),
however we know that the DOS and PES are not strictly comparable because of
the different photoelectric cross-section of the Ir-$d$ and O-$p$ states. In the following we have computed the PES 
within the so-called single-scatterer final-state approximation.~\cite{XPS1,XPS2} Here the photocurrent is a sum of
local (atomic-like) and partial ($l$-like) density of states weighted by the corresponding cross-sections.
In order to take into account the lifetime broadening which increase in proportion to the square of the binding energy within the Fermi
liquid theory, the spectrum has been broadened by a Lorentzian function. The broadening parameter of
the Lorentzian ($\Gamma$) is taken to be dependent with the photon energy (E),
having the form: $\Gamma = 0.04 + 0.03E^2$.   
The computed spectrum shows a very good agreement with the experimental data (see Fig.~\ref{dmft_spectra}) as far as the main peak position
and the bandwidth are concerned. 
Finally to investigate the importance of dynamical correlation over the static correlation 
on the electronic structure of IrO$_2$, we have carried out DMFT calculation including SOC and computed the photoemission 
spectrum. Fig.~\ref{dmft_spectra} displays the computed spectrum within the DMFT approach on top of the experimental spectra and
the spectra obtained within GGA+U+SOC approach. 
The two methods yield very similar spectra 
and hence we conclude that the static
correlation is sufficient to describe the nonmagnetic metallic phase of IrO$_2$. 
\section{Conclusions}
In conclusion, our detailed study of the electronic structure and spectroscopic properties of IrO$_2$ reveal that it is essentially an uncorrelated material with strong SOC.  As a result of strong SOC, the Ir $5d$ t$_{2g}$ states at the Fermi level largely retain the $J_{\rm eff}$ =1/2 character even in the metallic state. We show as a function of $U$, that IrO$_2$ transforms from a non-magnetic metal to an antiferromagnetic metal and eventually into an antiferromagnetic insulator, where the role of SOC is to reduce the critical values of $U$ necessary for the transitions. 
This shows that it is possible to tune the properties of IrO$_2$ by means of correlation effects. We have discussed ways to realize this experimentally, by means of tuning the ratio between Coulomb $U$ over bandwidth $W$, e.g. by alloying and negative chemical pressure.
The optical and photoemission spectra calculated including SOC are in good agreement with experiment, suggesting the importance of SOC to understand the electronic structure of IrO$_2$. 
\section{Acknowledgments}
The authors thank Department of Science and Technology, Government of India for financial support. Support from the EU - INDIA collaboration program MONAMI is acknowledged. O.E. acknowledges support from the Swedish Research Council (VR), the KAW foundation, eSSENCE, and the ERC (project 247062 - ASD). A.D. acknowledges financial support from VR, the Royal Swedish Academy of Sciences (KVA) and the Knut and Alice Wallenberg trust (KAW). 

\end{document}